\documentstyle[12pt]{article}
\newcommand{\be}{\begin{equation}}
\newcommand{\ee}{\end{equation}}
\newcommand{\bea}{\begin{eqnarray}}
\newcommand{\eea}{\end{eqnarray}}

\newcommand{\p}[1]{(\ref{#1})}
\newcommand{\lb}{\label}

\begin{document}
\begin{titlepage}

\vspace*{2.5cm}

\begin{center}
{\Large\bf New Model of Higher-Spin Particle}

\vspace{1.5cm}

{\large\bf Sergey Fedoruk}\,,\,\,\,{\large\bf Evgeny Ivanov}
{${}^\star $}
\vspace{1cm}

{\it Bogoliubov  Laboratory of Theoretical Physics, JINR,}\\
{\it 141980 Dubna, Moscow region, Russia} \\
\vspace{0.3cm}

{\tt fedoruk,eivanov@theor.jinr.ru}\\
\vspace{0.5cm} \setcounter{footnote}{0}

\end{center}
\vspace{0.5cm} \vskip 0.6truecm  \nopagebreak

\begin{abstract}
\noindent We elaborate on a new model of the higher-spin (HS) particle which makes manifest
the classical equivalence of the HS particle of the unfolded formulation and the HS
particle model with a bosonic counterpart of supersymmetry. Both these models emerge as two
different gauges of the new master system. Physical states of the master model are massless
HS multiplets described by complex HS fields which carry an extra $U(1)$ charge $q\,$. The
latter fully characterizes the given multiplet by fixing the minimal helicity as $q/2$. We
construct the twistorial formulation of the master model and discuss symmetries of the new
HS multiplets within its framework.
\end{abstract}

\bigskip
\vspace{1.5cm}

\vspace{3.5cm}

\begin{center}
{${\,}^\star \,$}{\it Talk given by E. Ivanov at the XII
International Conference on Symmetry Methods in Physics (SYMPHYS-XII),
Yerevan, Armenia, July 03 - 08,  2006.}
\end{center}

\newpage

\end{titlepage}

\section{Introduction}

\noindent Concise and suggestive formulations of higher spin (HS)
theory make use of (super)spaces with extra bosonic co-ordinates
(see e.g. \cite{FrVas}--\cite{Vas}). The simple and, at the same
time, powerful device for the analysis of the geometric structure
of such (super)spaces is provided by (super)particles propagating
in them.

The so-called unfolded formulation of the HS theory \cite{Vas} is
reproduced by quantizing the tensorial
particle~\cite{BandLuk}-\cite{BPST}, or the equivalent HS
particle~\cite{Vas}, in which tensorial coordinates were gauged
away. Such formulations possess $Sp(8)$ and $OSp(1|8)$ symmetries
(which are hidden in the second formulation). There also exists a
different formulation of the HS particle exhibiting invariance
under the even counterpart of supersymmetry~\cite{FedLuk}.
Quantization of this model produces HS multiplets with all
helicities, just as in the case of the unfolded HS particle. In
the next Section we briefly review the HS particle models of both
types.

Or basic aim in this contribution is to describe new (``master'') model of the HS
particle~\cite{FI-CQG}. It yields, on the classical level, both above--mentioned HS
particle models as its two appropriate gauges. In Sect.~3 we give a general description of
our model. The master HS particle propagates in a space which is parametrized, in addition
to the position four--vector, by two pairs of the commuting spinorial variables and an
extra complex scalar $\eta$. The set of the first class constraints includes the basic
unfolded constraints, the first class generalization of the spinor constraints and a scalar
constraint which generates local $U(1)$ transformations of the twistor--like spinor
variables and complex scalar co-ordinate.

In Sect.~4 we perform the quantization of the master system. In the relevant set of HS
equations for the wave function, the basic unfolded equation \cite{Vas} proves to be a
consequence of the quantum counterparts of the spinor constraints. There is also a scalar
$U(1)$ constraint which is an analog of the ``spin--shell'' constraint present in the model
of massless particle with fixed helicity~\cite{ES}, \cite{Town}. In our case the degree of
homogeneity of the HS field with respect to commuting twistor-like spinors is not fixed due
to the presence of complex scalar coordinate $\eta$ with non-zero $U(1)$ charge. The
external $U(1)$ charge $q$ of the HS wave function in extended space is defined as a degree
of homogeneity with respect to {\it both} spinor and scalar co-ordinates. Physical states
of the HS fields are massless particles with the helicities ranging from $\frac{q}{2}$ to
infinity. Besides the standard HS multiplet \cite{Vas} corresponding to the choice of
$q=0$, the considered setting implies the existence of new HS multiplets with non-zero
minimal helicities $\frac{q}{2}\,$, $q \neq 0\,$.

In Sect.~5 we construct a twistor formulation of the master system. The master HS particle
propagates in a space parameterized by a unit twistor and an additional complex scalar. We
construct a coordinate twistor transform relating different classical formulations of the
master system, as well as a field twistor transform which allows one to reconstruct the
space--time HS fields by the twistorial ``prepotential'' which solves the HS equations.
Using twistorial formulation of various HS multiplets, we find the HS algebras associated
with them.

In Sect.~6 we summarize our results.

\setcounter{equation}{0}

\section{A survey of the previously known HS particle \\ models}

\centerline{\bf HS particle related to the unfolded formulation}
\smallskip

\noindent There are two world--line interpretations of the
unfolded formulation of the HS field theory~\cite{Vas}. One of
them proceeds from the model of tensorial
superparticle~\cite{BandLuk,BandLukSor}. Actually, the latter can
be equivalently formulated either in a hyperspace containing a
ten--dimensional bosonic subspace alongside an extra commuting
Weyl spinor, or in superspace with the Grassmann spinor coordinate
(quantization of tensorial superparticle and links of it to an
unfolded formulation were also studied in~\cite{PST1,BPST}).

There also exists a version of the unfolded formulation which
makes no use of the tensorial coordinates at all~\cite{Vas}. In
the pure bosonic case the basic equation for the HS field $\Phi(x,
y,\bar y)$ \cite{Vas} reads
\begin{equation}\label{unfold-eq}
\left(\partial_{\alpha\dot\alpha} +i\frac{\partial}{\partial y^{\alpha}}
\frac{\partial}{\partial \bar y^{\dot\alpha}} \right) \Phi = 0\,,
\end{equation}
where $y^{\alpha}$ is a commuting Weyl spinor,
$\bar y^{\dot\alpha}=\overline{(y^{\alpha})}$.
Solution of the unfolded equation~(\ref{unfold-eq}) can be found,
assuming the polynomial dependence of the wave function on the
commuting spinors $y^{\alpha}$, $\bar y^{\dot\alpha}$
\begin{equation}\label{wf-tens}
\Phi(x, y, \bar y)  =\sum_{m=0}^{\infty}\sum_{n=0}^{\infty} y^{\alpha_1}\ldots y^{\alpha_m}
\bar y^{\dot\alpha_1}\ldots\bar y^{\dot\alpha_n} \varphi_{\alpha_1 \ldots \alpha_m
\dot\alpha_1 \ldots \dot\alpha_n}(x)\,.
\end{equation}
Independent space-time fields in the expansion of the general
field~(\ref{wf-tens}) are self--dual $\varphi_{\alpha_1 \ldots
\alpha_m}$ and anti--self--dual $\varphi_{\dot\alpha_1 \ldots
\dot\alpha_n}$ field strengths of all helicities (including the
half--integer ones). Basic unfolded equation~(\ref{unfold-eq})
leads to Klein--Gordon and Dirac equations for them. All other
component fields are expressed as $x$-derivatives of these basic
ones. Reality condition for the HS field  $\Phi(x, y,\bar y) =
\bar\Phi(x, y,\bar y)$ leads to the reality conditions
$\varphi_{\dot\alpha_1 \ldots \dot\alpha_n}=\bar\varphi_{\alpha_1
\ldots \alpha_m}$ for physical fields. Thus, the massless HS
multiplet described by the real HS field $\Phi(x, y,\bar y)$
contains all helicities, each helicity appearing only once.

A classical counterpart of this unfolded formulation is the particle system with the
action~\cite{Vas}
\begin{equation}\label{act-1}
S_{1}=\int d\tau \left( \lambda_\alpha \bar\lambda_{\dot\alpha} \dot x^{\dot\alpha\alpha} +
\lambda_\alpha\dot y^\alpha + \bar\lambda_{\dot\alpha}\dot{\bar y}^{\dot\alpha} \right).
\end{equation}
The spinors $\lambda_\alpha$, $\bar\lambda_{\dot\alpha}$ are
canonical momenta for $y^{\alpha}$, $\bar y^{\dot\alpha}\,$. The
first class constraints
\begin{equation}\label{P-res}
P_{\alpha\dot\alpha} - \lambda_\alpha
\bar\lambda_{\dot\alpha} \approx 0
\end{equation}
after quantization reproduce the unfolded equation~(\ref{unfold-eq}).

\bigskip

\centerline{\bf Model with an even counterpart of N=1 supersymmetry}

\smallskip

\noindent A different model of the massless HS particle was
proposed in~\cite{FedLuk}. The action resembles that of the usual
massless $N=1$ superparticle
\begin{equation}\label{act-bsusy}
S_{2}=\int d\tau \left( P_{\alpha\dot\alpha} \omega^{\dot\alpha\alpha} - e
P_{\alpha\dot\alpha}P^{\alpha\dot\alpha} \right),
\end{equation}
\begin{equation}\label{om}
\omega^{\dot\alpha\alpha} \equiv \dot x^{\dot\alpha\alpha} -i
{\bar\zeta}{}^{\dot\alpha}\dot{\zeta}{}^{\alpha}+i
\dot{\bar\zeta}{}^{\dot\alpha}{\zeta}^{\alpha}\,.
\end{equation}
However, the crucial difference from the superparticle case is
that the Weyl spinor $\zeta^\alpha$, $\bar\zeta^{\dot\alpha}=
(\overline{\zeta^\alpha})$, is commuting. The distinguishing
feature of this model is its manifest invariance under the even
counterpart of $4D$ supersymmetry
(SUSY)~\cite{FedZim,Lecht,FedLuk}
\begin{equation}\label{susy}
\delta x^{\dot\alpha\alpha} =
i(\bar\epsilon^{\dot\alpha}\zeta^\alpha
-\bar\zeta^{\dot\alpha}\epsilon^\alpha) \, ,\quad \delta
\zeta^\alpha = \epsilon^\alpha \, ,\quad \delta
\bar\zeta^{\dot\alpha}= \bar\epsilon^{\dot\alpha}\,,
\end{equation}
where $\epsilon^\alpha$ is a commuting Weyl spinor. The detailed
analysis of global symmetries of the model~(\ref{act-bsusy}) was
fulfilled in~\cite{FIL}. In particular, the even SUSY
translations~(\ref{susy}) are part of the $SU(3,2)$ group symmetry
of the system~(\ref{act-bsusy}).

The set of the Hamiltonian constraints of the system includes the
mass-shell constraint $
P_{\alpha\dot\alpha}P^{\alpha\dot\alpha}\approx 0 $ and the even
spinor constraints
\begin{equation}\label{cons-D}
D_\alpha\equiv \pi_\alpha
+iP_{\alpha\dot\alpha}\bar\zeta^{\dot\alpha}\approx
0\,,\qquad\qquad \bar D_{\dot\alpha}\equiv \bar\pi_{\dot\alpha}
-i\zeta^{\alpha}P_{\alpha\dot\alpha}\approx 0\,,
\end{equation}
where $\pi_\alpha$ and $\bar \pi_{\dot\alpha}$ are conjugate momenta
for $\zeta^\alpha$ and $\bar \zeta^{\dot\alpha}$.

The wave function of the particle model with
even ``supersymmetry''~(\ref{act-bsusy}) was obtained in~\cite{FedLuk} (see
also~\cite{FIL}, where a superextension of the
model~(\ref{act-bsusy}) was considered), and it is an even counterpart of
chiral $N=1$ superfield
\begin{equation}\label{even-sfield}
\Psi(x_{\!\scriptscriptstyle L}, \zeta)=
\sum_{n=0}^{\infty} \zeta^{\alpha_1}\ldots \zeta^{\alpha_n}
\psi_{\alpha_1 \ldots \alpha_n}
(x_{\!\scriptscriptstyle L})\,,
\end{equation}
where $x_{\!\scriptscriptstyle L}^{\dot\alpha\alpha} =
x^{\dot\alpha\alpha} +i
{\bar\zeta}{}^{\dot\alpha}{\zeta}{}^{\alpha}\,$. Besides the
chirality condition $\bar D_{\dot\alpha} \Psi =0$, this field is
subjected to the equations\footnote{Here and lower we use
following notation $\partial_\alpha
\equiv\frac{\partial}{\partial\zeta^{\alpha}}$,
$\bar\partial_{\dot\alpha} \equiv \frac{\partial}{\partial
\bar\zeta^{\dot\alpha}}$}
\begin{equation}\label{1-cl-con}
\partial_{\!\scriptscriptstyle L}^{\dot\alpha\alpha} \partial_{\alpha} \, \Psi =
0\,, \qquad
\partial_{\!\scriptscriptstyle L}^{\dot\alpha\alpha}
\partial_{{\!\scriptscriptstyle L}\,\alpha\dot\alpha}\, \Psi = 0\,,
\end{equation}
which are quantum counterparts of the first class constraints. Due to
eqs.~(\ref{1-cl-con}) the fields in the
expansion~(\ref{even-sfield}) are complex self--dual field
strengths of the massless particles of all helicities. As a
result, the spectrum contains all helicities, every non-zero
helicity appearing only once. In this picture the scalar field is
complex, as opposed to the real scalar field present in the HS
field~(\ref{wf-tens}) of the unfolded formulation.

\setcounter{equation}{0}

\section{Master HS particle model}

\noindent In a recent paper~\cite{FI-CQG} we proposed a new model of the even HS particle which
plays the role of the ``master system'' both for the particle model~(\ref{act-1})
corresponding to the unfolded formulation and for the model~(\ref{act-bsusy}) with the
explicit even ``supersymmetry''.

The master system involves the variables of both systems~(\ref{act-1}) and
(\ref{act-bsusy})
and also an additional complex scalar $\eta$
($\bar\eta= \eta^\dagger$).
The model is described by the following action
\begin{equation}\label{act-mast}
S=\int d\tau \left[\lambda_\alpha \bar\lambda_{\dot\alpha} \omega^{\dot\alpha\alpha} +
\lambda_\alpha\dot y^\alpha + \bar\lambda_{\dot\alpha} \dot{\bar y}^{\dot\alpha}
+i(\eta\dot{\bar\eta} - \dot\eta\bar\eta) - 2i\bar\eta\dot{\zeta}{}^{\alpha}\lambda_\alpha
+2i\eta \bar\lambda_{\dot\alpha} \dot{\bar\zeta}{}^{\dot\alpha} - l \left({\cal N} -
c\right) \right].
\end{equation}
The field $l$ acts as a Lagrange multiplier effecting the
constraint
\begin{equation}\label{H-mast}
{\cal N} -c \equiv i\,(y^\alpha\lambda_\alpha
-\bar\lambda_{\dot\alpha} {\bar y}^{\dot\alpha})- 2\eta\bar\eta -c
\approx 0\,,
\end{equation}
which generates, in the Hamiltonian formalism, local $U(1)$
transformations of the involved complex fields, except for the
fields $\zeta$, $\bar\zeta$.

The action~\p{act-mast} produces the following primary constraints
\begin{equation}\label{T-mast}
T_{\alpha\dot\alpha}\equiv P_{\alpha\dot\alpha} -\lambda_\alpha
\bar\lambda_{\dot\alpha}\approx 0\,,
\end{equation}
\begin{equation}\label{D-mast}
{\cal D}_\alpha\equiv D_\alpha +
2i\bar\eta\lambda_\alpha\approx 0\,,\qquad\qquad \bar{\cal D}_{\dot\alpha}\equiv
\bar{D}_{\dot\alpha} - 2i\eta\bar\lambda_{\dot\alpha}\approx 0
\end{equation}
and
\begin{equation}\label{g-mast}
g \equiv p_\eta +i\bar\eta \approx 0\,,\qquad \bar g \equiv \bar p_{\eta} -i\eta
\approx 0\,,
\end{equation}
where $D_\alpha$ and ${\bar D}_{\dot\alpha}$ are defined
in~(\ref{cons-D}) and $p_\eta$ and $\bar p_{\eta}$ are the
conjugate momenta for $\eta$ and $\bar \eta$, respectively. We at
once treat the variables $\lambda_\alpha$ and
$\bar\lambda_{\dot\alpha}$ as conjugate momenta for $y^\alpha$ and
$\bar y^{\dot\alpha}\,$.

The constraints~(\ref{g-mast}) possess a non-vanishing Poisson
bracket:
\begin{equation}\label{PB-g}
[g , \bar g ]_{{}_P}=2i\,.
\end{equation}
So they are second class and can be treated in the strong sense by
introducing Dirac brackets for them.  Then the complex scalar
$\eta$ and its complex conjugate form the canonical pair
\begin{equation}\label{br-eta}
[\eta, \bar\eta ]_{{}_D}={\textstyle\frac{i}{2}}\,.
\end{equation}

Let us now explain in which way the master system is gauge-equivalent to the HS
particle systems
\p{act-bsusy} and \p{act-1}.

The systems~(\ref{act-bsusy}) and (\ref{act-mast}) are (classically) equivalent
to each other in the common sector of their phase space. This sector is singled out
by choosing the
definite sign of the energy $P_0$ which is fixed due to the constraint~(\ref{T-mast}).
This equivalence becomes evident if one observes that the system~(\ref{act-mast})
can be interpreted
as the system~(\ref{act-bsusy}) in which the second class constraints are converted
into the first class ones by
introducing the new canonical pair $\eta$, $\bar\eta$.

To be more precise, we use the covariant conversion method firstly proposed
in~\cite{ES} for the case of usual superparticle.
To convert two second class constraints contained in the spinor
constraints~(\ref{cons-D}) into first class ones,
we introduce two additional degrees of freedom carried by the complex scalar field
$\eta$.
We also introduce a commuting Weyl spinor $\lambda_\alpha$ to ensure the Lorentz
covariance of
the new spinor constraints~(\ref{D-mast}).
The closure of the algebra of the new spinor constraints
\begin{equation}\label{PB-con}
[{\cal D}_\alpha , \bar{\cal D}_{\dot\alpha} ]_{{}_D}=2i \, T_{\alpha\dot\alpha}
\end{equation}
gives just the constraint~(\ref{T-mast}) resolving four-momentum in terms of the
spinor product.
This resolution is defined up to an arbitrary phase transformation of $\lambda_\alpha$
$\bar\lambda_{\dot\alpha}$. In order to ensure this $U(1)$ gauge invariance
in the full modified action, we are led to add the constraint~(\ref{H-mast}).

A heuristic argument why this equivalence should hold is that both
models, (\ref{act-bsusy}) and (\ref{act-mast}), have the same
number $n_{ph}={\bf 8}$ of the physical degrees of freedom. The
rigorous proof of the equivalence can be achieved by reducing both
systems to the physical degrees of freedom. Namely, choosing
light--cone gauge and following the gauge-fixing procedure as
in~\cite{ES}, \cite{Town}, we showed in~\cite{FI-CQG} that the
actions of the systems~(\ref{act-mast}) and~(\ref{act-bsusy})
written in terms of physical variables coincide with each other in
the sector with the definite sign of energy.

The world-line particle model~(\ref{act-1}) also follows from the
master model~(\ref{act-mast}) under a particular gauge choice. The
spinor constraints~(\ref{D-mast}) and the gauge-fixing condition
$\zeta^{\alpha} \approx 0$ together with its complex conjugate can
be used to eliminate the variables $\zeta^{\alpha}$, $\pi_\alpha$
and their complex conjugates. Then the constraint (\ref{H-mast})
can be used to gauge away the variable $\eta$. The
constraint~(\ref{H-mast}) is linear in $\rho\equiv \eta\bar\eta$
and generates arbitrary local transformations of
$\varphi-i\ln(\eta/\bar\eta)$. This constraint, together with the
gauge fixing condition $\chi \equiv \varphi - const \approx 0\,$, can
be used to eliminate the variables $\rho$, $\varphi$ at expense of
the variables $\lambda_\alpha$, $y^\alpha$,
$\bar\lambda_{\dot\alpha}$, ${\bar y}^{\dot\alpha}$. Since the
gauge fixing condition includes only $\varphi$, the brackets for
the remaining variables are not affected. As a result, we arrive at
the system~(\ref{act-1}).

\setcounter{equation}{0}

\section{First-quantized theory}

\noindent In the representation in which the operators $\hat
P_{\alpha\dot\alpha}$, $\hat\pi_{\alpha}$,
$\hat{\bar\pi}_{\dot\alpha}$, $\hat\lambda_{\alpha}$,
$\hat{\bar\lambda}_{\dot\alpha}$ and $\hat{\bar \eta}$ are
realized by differential operators, the equations on the wave
function $F^{(q)} (x, \zeta, \bar\zeta, y, \bar y, \eta)$ are
\begin{equation}\label{T-eq}
\left(\partial_{\alpha\dot\beta} +i\frac{\partial}{\partial
y^{\alpha}}
\frac{\partial}{\partial \bar y ^{\dot\beta}} \right) F^{(q)} = 0\,,
\end{equation}
\begin{equation}\label{D-eq}
\mbox{(a)}\;\;\left(D_{\alpha}+ \frac{\partial}{\partial \eta}
\frac{\partial}{\partial y^{\alpha}} \right) F^{(q)} = 0\,,
\quad\qquad \mbox{(b)}\;\;\left(\bar D_{\dot\alpha}- 2\eta
\frac{\partial}{\partial \bar y^{\dot\alpha}} \right)\, F^{(q)} =
0\,,
\end{equation}
\begin{equation}\label{H-eq}
\left(y^{\alpha}\frac{\partial}{\partial
y^{\alpha}} - \bar y^{\dot\alpha}\frac{\partial}{\partial
\bar y^{\dot\alpha}} - \eta\frac{\partial}{\partial
\eta} \right) F^{(q)} = q\,F^{(q)}\,.
\end{equation}
Here, the operators $ D_\alpha= -i (\partial_\alpha
+i\partial_{\alpha\dot\alpha} \bar\zeta^{\dot\alpha} )$ and ${\bar
D}_{\dot\alpha}= -i (\bar\partial_{\dot\alpha}
-i\zeta^{\alpha}\partial_{\alpha\dot\alpha})$ are quantum
counterparts of the ``covariant momenta''~(\ref{cons-D}).

The external $U(1)$ charge $q$ defined by~(\ref{H-eq}) is the quantum counterpart of
the ``classical'' constant $c$ present in the constraint~(\ref{H-mast}), now with
the ordering ambiguities taken into account. Eq.~(\ref{H-eq}) implies $U(1)$
covariance of the wave function
\begin{equation}\label{u1-inv}
F^{(q)} (x, \zeta, \bar\zeta, e^{i\varphi} y,
e^{-i\varphi} \bar y, e^{-i\varphi} \eta) =
e^{qi\varphi} F^{(q)} (x, \zeta, \bar\zeta, y, \bar y, \eta)\,.
\end{equation}
Requiring $F^{(q)}$ to be single-valued restricts $q$ to the
integer values. It is important to note that this $U(1)$
covariance just implies that any monomial of the charged
coordinates in the series expansion of the wave function $F^{(q)}$
has the same fixed charge $q$ and it does not entail any $U(1)$
transformation of the coefficient fields. In this respect $q$
resembles the ``harmonic $U(1)$ charge'' of the harmonic
superspace approach \cite{GIOS,GIKOS} and the operator in \p{H-eq}
is an analog of the $U(1)$ charge-counting operator $D^0$ in this
approach.

Though the system of eqs.~\p{T-eq}--\p{H-eq} involves the
Vasiliev-type unfolded vector equation~(\ref{T-eq}), the latter
follows from the spinorial eqs.~(\ref{D-eq}) as their
integrability condition. So the basic independent equations of the
system \p{T-eq}--\p{H-eq} are the bosonic spinorial equations
(\ref{D-eq}) and the $U(1)$ condition \p{H-eq}.

Let us solve eqs.~(\ref{D-eq})--(\ref{H-eq}). In the variables
$\zeta^{\alpha}$, $\bar\zeta^{\dot\alpha}$, $y^{\alpha}$, $\eta$
and
\begin{equation}\label{x,by-l}
x_{\!\scriptscriptstyle L}^{\dot\alpha\alpha} =
x^{\dot\alpha\alpha} + i\, \bar\zeta^{\dot\alpha}\zeta^{\alpha}\,,
\qquad \bar y_{\!\scriptscriptstyle L}^{\dot\alpha} = \bar
y^{\dot\alpha} + 2i\, \eta\,\bar\zeta^{\dot\alpha}\,.
\end{equation}
eqs.~(\ref{D-eq})--(\ref{H-eq}) take the form
\begin{equation}\label{D-eq-1}
\mbox{(a)}\;\;\left[ -i \left(\partial_{\alpha} +i
\frac{\partial}{\partial \eta} \frac{\partial}{\partial
y^{\alpha}}\right) + 2 \bar\zeta^{\dot\alpha}
\left(\partial_{{\!\scriptscriptstyle L}\,\alpha\dot\alpha}
+i\frac{\partial}{\partial y^{\alpha}} \frac{\partial}{\partial
\bar y_{\!\scriptscriptstyle L}^{\dot\alpha}} \right) \right]
F^{(q)} = 0\,,\qquad \mbox{(b)}\;\;\bar\partial_{\dot\alpha}
F^{(q)} =0\,,
\end{equation}
\begin{equation}\label{H-eq-1}
\left(y^{\alpha}\frac{\partial}{\partial
y^{\alpha}} - \bar y_{\!\scriptscriptstyle L}^{\dot\alpha}\frac{\partial}{\partial
\bar y_{\!\scriptscriptstyle L}^{\dot\alpha}} - \eta\frac{\partial}{\partial
\eta} \right) F^{(q)} = q \, F^{(q)}\,.
\end{equation}

Eq.~(\ref{D-eq-1}b) is the bosonic chirality condition stating
that $F^{(q)}$ does not depend on $\bar\zeta^{\dot\alpha}$ in the
new variables, $F^{(q)} = F^{(q)} (x_{\!\scriptscriptstyle L},
\zeta, y, \bar y_{\!\scriptscriptstyle L}, \eta)$. Then
eq.~(\ref{D-eq-1}a) amounts to the equations
\begin{equation}\label{D-eq-1a}
\left(\partial_{\alpha} +i \frac{\partial}{\partial
\eta} \frac{\partial}{\partial
y^{\alpha}}\right) F^{(q)} = 0
\end{equation}
and
\begin{equation}\label{T-eq-1}
\left(\partial_{{\!\scriptscriptstyle L}\,\alpha\dot\alpha}
+i\frac{\partial}{\partial y^{\alpha}} \frac{\partial}{\partial \bar
y_{\!\scriptscriptstyle L}^{\dot\alpha}} \right) F^{(q)} = 0\,.
\end{equation}

The solutions of eqs.~(\ref{H-eq-1}), (\ref{D-eq-1a}) and
(\ref{T-eq-1}) can be obtained in several equivalent ways.

\bigskip

\centerline{\bf The unfolded--type description}

\smallskip

\noindent By analogy with~(\ref{even-sfield}) we assume the
polynomial dependence of wave function on $\zeta^\alpha$
\begin{equation}\label{Phi-z-exp}
F^{(q)}(x_{\!\scriptscriptstyle L}, \zeta, y,
\bar y_{\!\scriptscriptstyle L}, \eta)=
\sum_{n=0}^{\infty} \zeta^{\alpha_1}\ldots \zeta^{\alpha_n}
\Phi^{(q)}_{\alpha_1 \ldots \alpha_n}
(x_{\!\scriptscriptstyle L}, y,
\bar y_{\!\scriptscriptstyle L}, \eta) \,.
\end{equation}
Eq.~(\ref{D-eq-1a}) expresses all the coefficients in this expansion
as derivatives of the first
coefficient $\Phi^{(q)}(x_{\!\scriptscriptstyle L},
 y, \bar y_{\!\scriptscriptstyle L}, \eta)$ satisfying
the two equations
\be
\mbox{(a)}\;\;\left(\partial_{{\!\scriptscriptstyle
L}\,\alpha\dot\alpha} +i\frac{\partial}{\partial y^{\alpha}}
\frac{\partial}{\partial \bar y_{\!\scriptscriptstyle
L}^{\dot\alpha}} \right) \Phi^{(q)} = 0\,, \quad
\mbox{(b)}\;\;\left(y^{\alpha}\frac{\partial}{\partial y^{\alpha}}
- \bar y_{\!\scriptscriptstyle
L}^{\dot\alpha}\frac{\partial}{\partial \bar
y_{\!\scriptscriptstyle L}^{\dot\alpha}} -
\eta\frac{\partial}{\partial \eta} \right) \Phi^{(q)} = q \,
\Phi^{(q)}\,. \lb{RedSys}
\ee

Like in the previous cases, we assume that $\Phi^{(q)}
(x_{\!\scriptscriptstyle L}, y, \bar y_{\!\scriptscriptstyle L},
\eta)$ has a non-singular polynomial expansion over the additional
coordinates. Then eq.~(\ref{RedSys}b) implies
\be
\Phi^{(q)}(x_{\!\scriptscriptstyle L}, y, \bar
y_{\!\scriptscriptstyle L}, \eta) = \sum_{k=0}^\infty \eta^k
\varphi^{(q+k)}(x_{\!\scriptscriptstyle L}, y, \bar
y_{\!\scriptscriptstyle L})\,, \lb{quplusk} \ee and \be
\left(y^{\alpha}\frac{\partial}{\partial y^{\alpha}} - \bar
y_{\!\scriptscriptstyle L}^{\dot\alpha}\frac{\partial}{\partial
\bar y_{\!\scriptscriptstyle L}^{\dot\alpha}}\right) \varphi^{(q +
k)} = (q + k)\, \varphi^{(q +k)} \,.\lb{RedCond}
\ee
The reduced
$U(1)$ condition \p{RedCond} fixes the $y, \bar y$ dependence of
the functions $\varphi^{(q+k)}$ as
\begin{equation}\label{field-gen}
\varphi^{(q+k)}(x_{\!\scriptscriptstyle L}, y,
\bar y_{\!\scriptscriptstyle L})= \!
\left\{ \!
\begin{array}{lr}
\sum\limits_{n=0}^{\infty} y^{\alpha_{1}}\ldots y^{\alpha_{q+k+n}}\bar
y_{\!\scriptscriptstyle L}^{\dot\beta_1}
\ldots \bar y_{\!\scriptscriptstyle L}^{\dot\beta_n}\,
\phi_{\alpha_{1} \ldots \alpha_{q+k+n}
\dot\beta_1 \ldots
\dot\beta_n}(x_{\!\scriptscriptstyle L})\,, &
\,\,\,\,\,(q+k)\!\geq 0, \\
\sum_{n=0}\limits^{\infty} y^{\alpha_{1}}\ldots y^{\alpha_{n}}\bar
y_{\!\scriptscriptstyle L}^{\dot\beta_1}
\ldots \bar y_{\!\scriptscriptstyle L}^{\dot\beta_{|q+k|+n}}\,
\phi_{\alpha_{1} \ldots \alpha_{n}
\dot\beta_1 \ldots
\dot\beta_{|q+k|+n}}(x_{\!\scriptscriptstyle L})\,, &
\,\,\,\,\, (q+k)\!< 0.\\
\end{array}
\right.
\end{equation}

It remains to find the restrictions imposed on the fields
$\varphi^{(q+k)}$ by the remaining unfolded equation
(\ref{RedSys}a).

It is easy to see that in the case $\underline{q = 0}$
eq.~(\ref{RedSys}a) expresses all the fields $\phi_{\alpha_1
\ldots \alpha_{k+n} \dot\beta_1 \ldots \dot\beta_n}$ with $n > 0$
in $\varphi^{(k)}$ as $x$-derivatives of the lowest component, the
self--dual field $\phi_{\alpha_1 \ldots \alpha_k}\,$. The latter
field satisfies Dirac and Klein--Gordon equations
\begin{equation}\label{eqs-D,KG}
\partial^{\dot\beta\alpha_1}\phi_{\alpha_1 \ldots
\alpha_k}  =0\,, \qquad \partial^{\dot\alpha\alpha}\partial_{\alpha\dot\alpha} \,\phi =0
\end{equation}
also as a consequence of the same eq.~(\ref{RedSys}a). Thus the
space of physical states of the model is spanned by the complex
self--dual field strengths $\phi_{\alpha_1 \ldots \alpha_k}$,
$k=0,1,2, \ldots\,$, of the massless particles of all integer and
half-integer helicities, and the case of $q=0$ basically amounts
to the standard HS multiplet of ref. \cite{Vas}.

Like in the $q=0$ case, for $\underline{q
> 0}$ eq.~(\ref{RedSys}a) expresses the fields $\phi_{\alpha_1
\ldots \alpha_{q+k+n} \dot\beta_1 \ldots \dot\beta_n}$ with $n >
0$ in~\p{field-gen} in terms of the $
\partial_{\alpha\dot\alpha}$-derivatives of the self-dual fields
$\phi_{\alpha_1 \ldots \alpha_{q+k}}$. Also, the same
eq.~(\ref{T-eq-1}) yields Dirac equations for the independent
fields
\begin{equation}\label{eqs-D}
\partial^{\dot\beta\alpha_1}\phi_{\alpha_1 \ldots
\alpha_{q+k}}  =0 \,, \qquad k=0,1,2,\ldots\,.
\end{equation}
Thus the space of physical states of the model is spanned by the
self--dual field strengths of the massless particles with
helicities $\frac{q}{2},\frac{q}{2}+\frac{1}{2},
\frac{q}{2}+1,\ldots\,$. We observe that the scalar field is
absent in the spectrum for non-zero positive $q\,$. The relevant
HS multiplet is fully characterized by the value of $q\,$.

For $\underline{q < 0}$ the
expansion~(\ref{quplusk}) can be conveniently rewritten as
\begin{eqnarray}\label{field-1-neg}
\Phi^{(q)}(x_{\!\scriptscriptstyle L}, y,
\bar y_{\!\scriptscriptstyle L}, \eta) &=&
\eta^{|q|}\tilde\Phi^{(0)}(x_{\!\scriptscriptstyle L}, y,
\bar y_{\!\scriptscriptstyle L}, \eta) +\\
&& + \sum_{k=1}^{|q|}
\sum_{m=0}^{\infty} \eta^{|q|-k}\, y^{\alpha_1}\ldots y^{\alpha_{m}}
\bar y_{\!\scriptscriptstyle L}^{\dot\beta_1}
\ldots \bar y_{\!\scriptscriptstyle L}^{\dot\beta_{m+k}}\,
\phi_{\alpha_1 \ldots \alpha_{m}
\dot\beta_1 \ldots
\dot\beta_{m+k}}(x_{\!\scriptscriptstyle L})\,.
\nonumber
\end{eqnarray}
Thus in this case we deal with the HS field $\tilde\Phi^{(0)}$
having the same helicity contents as the $q=0$ multiplet and an
additional term involving the space--time fields $\phi_{\alpha_1
\ldots \alpha_{m}\dot\beta_1 \ldots \dot\beta_{m+k}}\,$. The set
of independent fields in this expansion consists of self--dual
fields $\phi_{\alpha_1 \ldots \alpha_k}$, $k=0,1,2, \ldots$
present in the HS field $\Phi^{(0)}$ and anti--self--dual fields
$\phi_{\dot\beta_1 \ldots \dot\beta_{k}}$, $k=1,\ldots, |q|$.
Eq.~(\ref{RedSys}a) expresses all other space--time fields as
space--time derivatives of these basic ones. Eq.~(\ref{RedSys}a)
also implies the Dirac and Klein--Gordon equations for the basic
fields. Thus, for $q< 0$ physical fields in the spectrum describe
massless particles with all positive helicities starting from the
zero one, and also a finite number of massless states with
negative helicities $-\frac{1}{2}, -1, \ldots, -\frac{|q|}{2}$.
This HS multiplet can be naturally called ``helicity-flip''
multiplet. Note that, being considered together with its
conjugate, this multiplet reveals a partial doubling of fields
with a given helicity, the phenomenon which is absent in the
previous two cases.

\bigskip

\centerline{\bf The description with explicit even SUSY}

\smallskip

\noindent We start with the case $q=0$ and consider at first the
expansion of wave function with respect to the coordinates $y$,
$\bar y_{\!\scriptscriptstyle L}$ and $\eta$
\begin{equation}\label{Phi-left-1}
F^{(0)} (x_{\!\scriptscriptstyle L}, \zeta, y, \bar y_{\!\scriptscriptstyle L},
\eta) = \sum_{k=0}^{\infty}
\sum_{n=0}^{\infty} \eta^{k}\, y^{\alpha_1}\ldots y^{\alpha_{k+n}}
\bar y_{\!\scriptscriptstyle L}^{\dot\alpha_1}
\ldots \bar y_{\!\scriptscriptstyle L}^{\dot\alpha_n}\,
\Psi_{\alpha_1 \ldots \alpha_{k+n}
\dot\alpha_1 \ldots
\dot\alpha_n}(x_{\!\scriptscriptstyle L}, \zeta) \,,
\end{equation}
where we have already taken care of eq.~(\ref{H-eq-1}). It remains
to take into account eqs.~(\ref{D-eq-1a}) and~(\ref{T-eq-1}). They
express all the coefficient fields in~\p{Phi-left-1} as
derivatives of the lowest coefficient
$\Psi(x_{\!\scriptscriptstyle L}, \zeta)\,$ which is exactly the
chiral HS field~(\ref{even-sfield}) of ref.~\cite{FedLuk}. It
comprises the same irreducible HS $q=0$ multiplet as $\Phi^{(0)}
(x_{\!\scriptscriptstyle L}, y, \bar{y}_{\!\scriptscriptstyle L},
\eta)\,$. The possibility to describe the same multiplet in two
equivalent ways is just the quantum manifestation of the
equivalence between the HS particles \p{act-1} and \p{act-bsusy}
which both follow from the master HS particle. In a sense, the
description by the field $\Psi(x_{\!\scriptscriptstyle L},\zeta)$
is more economical since this quantity contains, in its
$\zeta^\alpha$ expansion, just the independent space-time
self-dual fields (anti-self-dual fields are collected by the
complex--conjugated function $\bar\Psi (x_{\!\scriptscriptstyle
R}, \bar\zeta)\,$).

The $q > 0$ counterpart of the $q=0$ expansion is
\begin{equation}\label{field-q-pos}
F^{(q)}(x_{\!\scriptscriptstyle L}, \zeta, y,
\bar y_{\!\scriptscriptstyle L}, \eta)= \sum_{k=0}^{\infty}
\sum_{n=0}^{\infty} \eta^{k}\, y^{\alpha_1}\ldots y^{\alpha_{q+k+n}}
\bar y_{\!\scriptscriptstyle L}^{\dot\beta_1}
\ldots \bar y_{\!\scriptscriptstyle L}^{\dot\beta_n}\,
\Psi_{\alpha_1 \ldots \alpha_{q+k+n}
\dot\beta_1 \ldots
\dot\beta_n}(x_{\!\scriptscriptstyle L}, \zeta)\,,
\end{equation}
where we have already taken into account the $U(1)$ condition
(\ref{H-eq-1}). Then  eqs.~(\ref{D-eq-1a}) and~(\ref{T-eq-1}) lead
to the expressions for all component fields in terms of the even
counterpart of $N=1$ chiral field with external indices
$\Psi_{\alpha_1 \ldots \alpha_q} (x_{\!\scriptscriptstyle L},
\zeta)\,$. As a consequence of eqs.~(\ref{D-eq-1a})
and~(\ref{T-eq-1}), this field is subjected to the
equations
\begin{equation}\label{Su-eq}
\partial_{\!\scriptscriptstyle L}^{\dot\alpha\alpha} \partial_{\alpha} \, F^{(q)} =
0\,, \qquad
\partial_{\!\scriptscriptstyle L}^{\dot\alpha\alpha}
\partial_{{\!\scriptscriptstyle L}\,\alpha\dot\alpha}\, F^{(q)} = 0\,,
\end{equation}
\begin{equation}\label{Dir-q}
\partial^{\alpha_1}\Psi_{\alpha_1 \ldots \alpha_q} =0\,, \qquad
\partial_{\!\scriptscriptstyle L}^{\dot\alpha\alpha_1}
\Psi_{\alpha_1 \ldots \alpha_q} =0  \,.
\end{equation}
Expanding the field $\Psi_{\alpha_1 \ldots \alpha_q}$ in powers of $\zeta_\alpha$
\begin{equation}\label{Phi-exp-q}
\Psi_{\alpha_1 \ldots \alpha_q}
(x_{\!\scriptscriptstyle L}, \zeta)
= \sum_{n=0}^{\infty} \zeta^{\beta_1}\ldots \zeta^{\beta_n}
\psi_{\alpha_1 \ldots \alpha_q \beta_1 \ldots \beta_n}
(x_{\!\scriptscriptstyle L})\,,
\end{equation}
we observe that all component fields in this expansion are totally
symmetric in the spinor indices due to the first equation
in~(\ref{Dir-q}). As a consequence of eqs.~(\ref{Su-eq}) and the
second equation in~(\ref{Dir-q}), all component fields satisfy
Dirac equation. Therefore, the HS field \p{Phi-exp-q} describes
the same physical spectrum as in (\ref{eqs-D}).
\setcounter{equation}{0}

\section{Twistorial formulation of the master HS particle}

\centerline{\bf Action of the master HS particle in twistorial
formulation}

\smallskip

\noindent The twistorial formulation of the master HS
particle~(\ref{act-mast}) was constructed in \cite{FI-CQG} and is
a HS generalization of the well--known twistor formulation of
massless particles with fixed helicities. It is described by two
Weyl spinors $\lambda_{\alpha}$ and $\bar\mu^{\dot\alpha}$ and a
complex scalar $\xi$ which are introduced by the following twistor
transform
\begin{equation}\label{mu}
\mu^{\alpha} = y^{\alpha} +\bar\lambda_{\dot\beta}
(x^{\dot\beta\alpha} -i\bar\zeta^{\dot\beta}\zeta^\alpha ) -
2i\,\bar\eta\, \zeta^\alpha \, , \qquad \bar\mu^{\dot\alpha} =\bar
y^{\dot\alpha}+ (x^{\dot\alpha\beta}
+i\bar\zeta^{\dot\alpha}\zeta^\beta )\lambda_{\beta} + 2i\,\eta\,
\bar\zeta^{\dot\alpha}\,,
\end{equation}
\begin{equation}\label{xi}
\xi=\eta +\zeta^\beta \lambda_\beta \, ,\qquad \bar\xi =\bar\eta
+\bar\lambda_{\dot\beta} \bar\zeta^{\dot\beta}\,.
\end{equation}
The spinors $\lambda_{\alpha}$, $\bar\mu^{\dot\alpha}$ are the
components of the twistor ($SU(2,2)$ spinor) $Z_a =
(\lambda_{\alpha}, \bar\mu^{\dot\alpha})$, $a=1,...,4\,$, in the
basis where it splits into irreps of the Lorentz group $SL(2,C)$
and dilatations $SO(1,1)$.

Up to some boundary terms, the action~(\ref{act-mast})
takes the following form in the twistorial variables
\begin{equation}\label{act-twist-2}
S^{HS-tw.}=\int d\tau \left[\lambda_\alpha\dot \mu^\alpha
+\bar\lambda_{\dot\alpha}
\dot{\bar \mu}^{\dot\alpha}+i(\xi\dot{\bar\xi} -
\dot\xi\bar\xi) - l \, \left(U - c\right)\right].
\end{equation}
Here, the $U(1)$ constraint
\begin{equation}\label{U-cons}
U - c\equiv i(\mu^\alpha\lambda_\alpha
-\bar\lambda_{\dot\alpha} {\bar\mu}^{\dot\alpha})- 2\xi\bar\xi -c
\approx 0
\end{equation}
is the condition (\ref{H-mast}) rewritten in the twistorial variables~(\ref{mu}),
(\ref{xi}).

The twistorial formulation~(\ref{act-twist-2}) of the HS particle
reproduces the twistorial HS particle which was considered
in~\cite{BandLuk,BandLukSor}. Due to the
constraint~(\ref{U-cons}), one can gauge away the variables $\xi$,
$\bar\xi$ as in Section III. As a result, we obtain the system
described by the action of ref. \cite{BandLuk,BandLukSor}
\begin{equation}\label{act-twist-1}
S_1^{HS-tw.}=\int d\tau \left(\lambda_\alpha\dot \mu^\alpha
+\bar\lambda_{\dot\alpha}
\dot{\bar \mu}^{\dot\alpha} \right).
\end{equation}
The twistor system~(\ref{act-twist-1}) can be also obtained
directly from the system~(\ref{act-1}) via the following standard
twistor transform $\mu^{\alpha} = y^{\alpha}
+\bar\lambda_{\dot\beta} x^{\dot\beta\alpha}$ ,
$\bar\mu^{\dot\alpha} =\bar y^{\dot\alpha}+
x^{\dot\alpha\beta}\lambda_{\beta}$. This provides one more proof
of the equivalence of the systems~(\ref{act-1})
and~(\ref{act-mast}).

\bigskip

\centerline{\bf Quantization in twistor formulation: twistor
transform for HS fields}

\smallskip

\noindent Quantization in the twistorial formulation~(\ref{act-twist-2}) gives rise to the
same results as in Section~4. In the ``twistorial representation'', when $\lambda_\alpha$,
$\bar\mu^{\dot\alpha}$ and $\xi$ are diagonal, the twistorial wave function
$G^{(q-2)}(\lambda, \bar\mu, \xi)$ satisfies the quantum counterpart of the
constraints~(\ref{U-cons})
\begin{equation}\label{H-eq-tw}
\left(-\lambda_{\alpha}\frac{\partial}{\partial
\lambda_{\alpha}} - \bar \mu^{\dot\alpha}\frac{\partial}{\partial
\bar \mu^{\dot\alpha}} - \xi\frac{\partial}{\partial
\xi} \right) G^{(q-2)} = (q-2)\,G^{(q-2)}\,.
\end{equation}

The wave function in the space--time description is derived by
analogy with the standard twistor approach~\cite{PenMac}. One
substitutes the incidence conditions~(\ref{mu}) and~(\ref{xi}) for
the variables $\bar\mu^{\dot\alpha}$ and $\xi$ in the  twistorial
wave function and performs a Fourier-type integral transformation
from $\lambda_\alpha$ to its canonically conjugated variable
$y_\alpha$
\begin{equation}\label{int-tr}
F^{(q)} (x_{\!\scriptscriptstyle L}, \zeta, y,
\bar y_{\!\scriptscriptstyle L}, \eta) =
\int d^2\lambda\,  e^{i y^\alpha \lambda_\alpha}
G^{(q-2)}(\lambda_\alpha, \bar y_{\!\scriptscriptstyle L}^{\dot\alpha}+
x_{\!\scriptscriptstyle L}^{\dot\alpha\beta} \lambda_{\beta} ,
\eta +\zeta^\beta \lambda_\beta)\,.
\end{equation}
Note that the variables $x_{\!\scriptscriptstyle L}$ and
$\bar y_{\!\scriptscriptstyle L}$ defined in~(\ref{x,by-l})
already appeared in the twistor transform~(\ref{mu}) for $\bar\mu$. The
integrand in~(\ref{int-tr}) includes the Fourier exponent in contrast to the Penrose
integral
transform~\cite{PenMac}.

Using the particular dependence of the twistorial field
$G^{(q-2)}\,$ on the involved co-ordinates, it is easy to check
that the field $F^{(q)}$ defined by~(\ref{int-tr}) automatically
satisfies eqs.~(\ref{H-eq-1}), (\ref{D-eq-1a}) and (\ref{T-eq-1}).
Thus, the twistorial formulation solves eqs.~(\ref{H-eq-1}),
(\ref{D-eq-1a}) and (\ref{T-eq-1}) in terms of the unconstrained
``prepotential'' $G^{(q-2)}(\lambda, \bar\mu, \xi)$.

\bigskip

\centerline{\bf Symmetries of HS multiplets}

\smallskip

\noindent The symmetry analysis is direct in the twistor formulation in which
different HS multiplets labelled by the $U(1)$ charge $q$ are specified by the single
eq.~(\ref{H-eq-tw}). We can find the symmetries following the techniques exploited
in ref. \cite{Vas}. The HS fields depend on the twistor variables $Z_a = (\lambda_{\alpha}, \bar\mu^{\dot\alpha})$,
$a=1,...,4\,$, and complex scalar $\xi$. The symmetry generators are products of the
$Z_a$, $\xi$ monomials of arbitrary degree and those of the derivatives $\frac{\partial}{\partial
Z_{a}}$, $\frac{\partial}{\partial \xi}$, such that they preserve eq.~(\ref{H-eq-tw}) rewritten
in the form
\begin{equation}\label{U-eq-tw}
\left(\hat U - \hat q\right) G^{(\hat q)} = 0\,,
\end{equation}
where
\begin{equation}\label{hU}
\hat U \equiv -Z_{a}\frac{\partial}{\partial Z_{a}} - \xi\frac{\partial}{\partial \xi} \,,
\qquad \hat q\equiv q-2\,.
\end{equation}

Let us consider the generators
\begin{equation}\label{gener-gen}
T^{\{ {\cal N}\}} (Z,\xi) \equiv T^{\langle N;K\rangle} (Z,\xi) = T^{\langle N\rangle} (Z)
\cdot T^{\langle K\rangle} (\xi)\,,\qquad {\cal N}=N+K\,,
\end{equation}
where the quantities
\begin{equation}\label{gener-Z}
T^{\langle N\rangle} (Z) \equiv T^{(n,m)} (Z) \equiv T_{a_1\cdots a_n}^{b_1 \cdots b_m} (Z)
= Z_{a_1}\cdots Z_{a_n} \frac{\partial}{\partial Z_{b_1}}\cdots \frac{\partial}{\partial
Z_{b_m}}\,, \qquad N=n+m
\end{equation}
act in the twistor sector, whereas the quantities \footnote{For definiteness, we use the
$\hat {\cal Q} \hat {\cal P}$--ordering with respect to $Z_a$, $\xi$ and their
``momenta''.}
\begin{equation}\label{gener-xi}
T^{\langle K\rangle} (\xi) \equiv T^{(k,\,l)} (\xi) = \xi^k \frac{\partial^{\,l}}{\partial
\xi^{\,l}}\,, \qquad K=k+l
\end{equation}
act on the scalar $\xi$. The generators~(\ref{gener-gen}) form an infinite--dimensional
algebra (modulo some coefficients)
\begin{equation}\label{inf-dim}
\left[ T^{\{ {\cal N}\}} , T^{\{ {\cal M}\}} \right] =\sum_{{\cal L}=0}^{{\cal N}+{\cal
M}-2} T^{\{ {\cal L}\}}\,.
\end{equation}

The symmetry algebra of the physical states described by HS field $G^{(\hat q)}$ is formed
by the generators $F^{(n,\,m ;\,k,\,l)}$ from~(\ref{gener-gen}) commuting with the
operator~(\ref{hU}):
\begin{equation}\label{FU}
\left[ F^{(n,\,m ;\,k,\,l)} , \hat U \right] =0\,.
\end{equation}
Using
$
\left[ T^{(n,\,m ;\,k,\,l)} , \hat U \right] =\left(n+k-m-l\right) T^{(n,\,m ;\,k,\,l)}
$
we find that $F^{(n,\,m ;\,k,\,l)}$ are the generators~(\ref{gener-gen}) with $n+k=m+l$. We
denote this infinite--dimensional algebra by $hsc(3,2)$ (with {\it hsc} for {\it higher
spin conformal}) to emphasize the presence of the finite-dimensional subalgebra $u(3,2)$
with the generators
\begin{equation}\label{u32}
F_a{}^b = Z_{a}\frac{\partial}{\partial Z_{b}}\,, \qquad F_a=Z_{a}\frac{\partial}{\partial
\xi}\,, \qquad F^b =\xi\frac{\partial}{\partial Z_{b}} \,, \qquad F
=\xi\frac{\partial}{\partial \xi}\,.
\end{equation}
The generators $F_a{}^b$ form $u(2,2)$ algebra which is an extension of the conformal
algebra $su(2,2)$ by the generator of phase transformations~(\ref{hU}) of twistor. The
operators $F_a$ and $F^b$ generate bosonic supersymmetry translations and bosonic
superconformal boosts~\cite{FIL}. The subalgebra $hsc(2,2)$ formed by the pure twistorial
generators~(\ref{gener-Z}) with $n=m$ produces higher spin algebras explored
in~\cite{FL,SeS,Vas}.

The algebra $hsc(3,2)$ is not simple since it contains ideals $I_{\hat q}$ spanned by the
elements of the form
\begin{equation}\label{H}
H^{(n,\,m ;\,k,\,l)} = \left( \hat U - \hat q\right) F^{(n,\,m ;\,k,\,l)} \,.
\end{equation}
However, the operators~(\ref{H}) become trivial on the HS multiplet $G^{(\hat q)}$.
Therefore $G^{(\hat q)}$ is associated with the quotient algebra $hsc_{\hat q}(3,2)=
hsc(3,2)/I_{\hat q}$. Different $\hat q$ correspond to different quotients of $hsc(3,2)$
associated with different HS multiplets. So the quotient algebras $hsc_{\hat q}(3,2)$ play
the role of ``primary'' symmetry algebras for the HS multiplets $G^{(\hat q)}\,$ considered
as their moduli. Here we do not discuss other symmetries of HS multiplets which are hidden
in the twistorial formulation (see a comment in~\cite{FI-CQG}).

\setcounter{equation}{0}

\section{Summary}

\noindent In this paper we have described a new HS particle model. This model is unifying
(``master'') for the unfolded HS particle and the HS particle with the even
``supersymmetry'', and it yields both these models upon choosing the appropriate gauges.

After quantization, the unfolded formulation of HS fields and their formulation with
the explicit bosonic ``supersymmetry'' are equivalent to each other as
they correspond to different ways of solving the same master
system of HS equations.
One of the novel features of this system is that
the infinite towers of
higher spins in the quantum spectrum are
accommodated by some holomorphic functions
depending on a new scalar complex bosonic variable $\eta\,$.
These functions are
characterized by the ``external'' $U(1)$ charge number $q$ which
fully specifies the corresponding infinite-dimensional multiplet
of spins. The HS fields respect a local $U(1)$ symmetry which is
similar to the $U(1)$ covariance of the harmonic approach
\cite{GIOS,GIKOS}. Crucial for maintaining this covariance is the
holomorphic dependence of the HS wave function on the complex
coordinate $\eta\,$.

Depending on their external $U(1)$ charge $q$, the HS fields in the extended space
accommodate different HS multiplets of ordinary $4D$ fields. The all-helicity HS multiplet
of the unfolded formulation (with a complex scalar field) is recovered as the $q=0$
multiplet and its conjugate. Also, some new HS multiplets with $q \neq 0$ emerge. For $q>0$
they are spanned by the self--dual field strengths of growing positive helicities, starting
from $\frac{q}{2}$. The $q<0$ multiplets show up an interesting ``spin--flip'' feature:
they include self-dual fields of all positive helicities, as well as a finite number of
anti--self--dual fields with negative helicities.  The complementary helicities are
accommodated by the complex conjugate wave functions.

Finally, let us note that the master model presented here
can play an important role in supersymmetric extensions of HS theory
(i.e. those with standard ``odd'' supersymmetries).
As shown in~\cite{FIL}, the supersymmetric $N=1$ HS theories constructed
as extensions of theories with bosonic ``supersymmetry'' are the
simplest HS theories respecting the fundamental notion of
chirality which underlies the geometric approach to the ordinary $N=1$
supergravity~\cite{OS}. The existence of a chiral limit seems to
be crucial for any satisfactory HS superfield theory including HS
generalizations of $N=1$ supergravity~\cite{BPST,IvLuk}.
Odd--supersymmetric extensions of the master model could provide further
insights into these and related issues.

\bigskip
\bigskip
%\section*{Acknowledgments}
\noindent {\bf Acknowledgments.} One of us (E.I.) would like to thank the Organizers of the
XII {\it International Conference on Symmetry Methods in Physics} for the kind hospitality
in Yerevan and a very pleasant friendly atmosphere created there. We wish to thank J.A.~de
Azc\'arraga, I.~Bandos, J.~Lukierski, D.~Sorokin and M.~Vasiliev for interest in the work
and useful remarks. This work was partially supported by the RFBR grant 06-02-16684 and the
grant INTAS-05-7928.

\end{document}